\documentclass[preprints,article,accept,moreauthors,pdftex]{Definitions/mdpi} 

\usepackage{bm}
\usepackage{bbm}
\usepackage{listings}
\usepackage{subfig }
\usepackage{amsmath}
\usepackage{amssymb}
\usepackage{amsthm}
\usepackage{booktabs}
\usepackage{hyperref}
\usepackage{erewhon} 
\usepackage[sf,scale=1]{libertine}
\usepackage{fourier}
\usepackage[T1]{fontenc}
\usepackage{algorithm}
\usepackage{algorithmic}

\usepackage{tikz}
\usetikzlibrary{arrows,shapes, positioning,fit,calc, decorations.pathmorphing,matrix}
\tikzstyle{inputNode}=[draw=none,rectangle, rounded corners, minimum size=10pt,inner sep=2pt,fill=green!50]
\tikzstyle{stateTransition}=[->, decorate, thick, shorten <= 2pt, shorten >= 2pt]
\tikzstyle{hiddenA}=[draw,rectangle, text width=360pt,	minimum width=360pt, minimum height=12pt,inner sep=2pt,fill=red!10]
\tikzstyle{hiddenB}=[draw,rectangle, text width=360pt,	minimum width=360pt, minimum height=12pt,inner sep=2pt,fill=yellow!20]
\tikzstyle{hiddenC}=[draw,rectangle, text width=360pt,	minimum width=360pt, minimum height=12pt,inner sep=2pt,fill=magenta!15]
\tikzstyle{output}=[draw=none,fill=blue!30,rectangle, rounded corners,minimum size=10pt,inner sep=1pt]

\hypersetup{colorlinks,citecolor=teal}

\firstpage{1} 
\pubvolume{xx}
\issuenum{1}
\articlenumber{1}
\pubyear{2022}
\copyrightyear{2022}
\history{}

\newcommand*\diff{\mathop{}\!\kern0pt\mathrm{d}}

\Title{Accurate and consistent calculation of the mean and variance in Monte-Carlo simulations}
\Author{Jherek Healy}
\AuthorNames{Jherek Healy}
\address{}
\corres{Correspondence: jherekhealy@protonmail.com}
\abstract{In parallelized Monte-Carlo simulations, the order of summation is not always the same. When the mean is calculated in running fashion, this may create an artificial randomness in results which ought to be reproducible. This note takes a look at the problem and proposes to combine the running mean and variance algorithm with an accurate and robust summing algorithm in order to increase the accuracy and robustness of the Monte-Carlo estimates.}

\begin{document}

\section{Introduction}
The straightforward and naive ways of calculating the mean and variance of the samples of a Monte-Carlo simulation may introduce artificial errors in the results.

Monte-Carlo simulations are inherently parallelisable. It is possible and often desirable to run such a parallel simulation using a consistent set of random numbers, such that the results of the simulation are reproducible. However, the operating system scheduling of the simulation on the multiple cores or CPUs implies some randomness in the order of the completed batches of results, due to the varying load of the CPUs. When the mean and variance of the Monte-Carlo simulation are calculated in running manner, the results obtained may differ due to the order of the terms in the summations involved, because of limits of accuracy imposed by the IEEE 64-bit floating-point standard. The differences may appear minor, but are amplified in finite-difference based sensitivities based on the mean, reusing the same random numbers (which is a standard practice). The use of a more robust algorithm will reduce, if not suppress, the variation of the mean and sensitivities due to the order of the samples. We assume, of course, that the set of random numbers used by the Monte-Carlo simulation is independent of the number of CPUs involved such that the Monte-Carlo samples are always reproducible in theory.

On one hand, the problem of calculating accurately the mean and variance has been studied in details in \citep{youngs1971some,ling1974comparison,chan1979computing,chan1983algorithms}. 
The recommendation of \citet{chan1979computing} is to use either
\begin{subequations}
\begin{equation}\label{eqn:chan_mean}
	M_{k} = M_{k-1} + \frac{x_{k}-M_{k-1}}{k}\,,
\end{equation}
or
\begin{equation}\label{eqn:naive_mean}
	S_{k} = S_{k-1} + x_{k}\,,\quad M_{k} = \frac{S_{k}}{k}\,,
\end{equation}
\end{subequations}
  as single pass algorithm for the mean $ M_{k}$, where $x_{k}$ is a new datum, along with
  \begin{subequations}
  \begin{equation}\label{eqn:chan_var}
	T_{k} = T_{k-1} + (k-1)\frac{(x_{k}-M_{k-1})^2}{k}\,,
\end{equation}  	
for the  variance $V=T_{k}/k$. In contrast, a naive implementation of the variance would consist in the iteration
  \begin{equation}\label{eqn:naive_var}
	T_{k} = T_{k-1} + x_{k}^2\,,
\end{equation}
along with $V = T_{k}/k-M_{k}^2$ for the last item.
The naive implementation suffers from catastrophic cancellation. A simple remedy is to add a shift, as the variance of $X-K$ is the same as the variance of $X$ where $K$ is a constant and $X$ a random variable. Using the first observation is often a good enough guess for $K$ \citep{chan1983algorithms}. This leads to the iteration
  \begin{equation}\label{eqn:naive_var_shifted}
	T_{k} = T_{k-1} + (x_{k}-K)^2\,,
\end{equation}
along with $V = T_{k}/k-(M_{k}-K)^2$ for the last item.
\end{subequations}
 
Double-pass algorithms are in general more robust and accurate for the sample variance, but require to store all the numbers in memory, which may not always be practical. Hence, this note focuses on the single-pass algorithms.

On the other hand, \citet{kahan1965further,klein2006generalized} propose to add some correction terms for the problem of calculating a running sum with increased accuracy.  \citet{richard1997adaptive} describes an algorithm which guarantees the reproducibility of the sum, but is much slower and thus not so practical, and \citet{neal2015fast} gives a fast exact summation algorithm, but it is much more complex to implement and only valid for 64-bit floating point numbers.

The two problems have been explored almost always in isolation, with the exception of \citet{schubert2018numerically}, which shows that including the summation correction increases the accuracy of the algorithm for the variance of \citet{youngs1971some}, and \citet{tian2012scalable}, which combines the Kahan summation with Equation \ref{eqn:chan_mean} for the calculation of the mean. This note intends to clarify what can be gained by combining the two approaches and including the correction terms to the calculation of $M_{k+1}, S_{k+1}, T_{k+1}$ above. We evaluate the different algorithms on theoretical examples as well as on concrete Monte-Carlo simulations.

\section{More threads, same numbers}\label{sec:more_threads}
The main application of this paper is to have reproducible Monte-Carlo results, independently of the number of concurrent threads (or CPUs) used. A typical Monte-Carlo simulation consists in generating $N D$ random numbers where $N$ is the number of Monte-Carlo paths, $D$ the number of dimensions of the problem (the number of independent random numbers needed to build one path), to build the paths according to some model, and to evaluate some function (a payoff in financial terms) on each path. 
A path is built using $M$ time-steps, and we have $D = dM$ where $d$ is the number of dimensions necessary\footnote{We assume here it to be constant, which is necessary for the use of low discrepancy sequences. The second approach, described in this section, is more general and allows for a variable number of dimensions. The other approaches would presuppose the random number generator to be able to generate directly any distribution and the random samples $u_{i,j}$ are not necessarily uniform.} to compute the path at the next time-step.
We would like to use the same numbers in and build the same paths, regardless of the number of threads. A single-threaded simulation is expected to produce exactly the same results. There are several ways to accomplish this.

A first approach is to store all the numbers in memory, and build each path in parallel, typically using a thread pool to limit the number of threads running at the same time. The $k$-th path would then refer to the specific indices $(k-1)  D + 1$ to $k  D$ in the array of random numbers, which may be accessed concurrently. It has two drawbacks: the size of this array may be too large to store in memory and the random number generation is not parallelized.

\begin{figure}[h!]
\begin{tikzpicture}
	\node[hiddenA] (thread1) at (-10,-0) {Thread A};
	\node[above=10pt of thread1.west] at (-13.5,0) {$t_1$};
	\node at (-11, 0.5) {$t_2$};
	\node (dots) at (-9.4,0.5)  {$\cdots$};
	\node at (-7.5, 0.5) {$t_M$};
	
	\node[inputNode, right=50pt of thread1.west] (t1) {$\tiny u_{1,1} , u_{1,2}, ..., u_{1,d}$};
	\node[inputNode, right=2pt of t1] (t2) {$\tiny u_{1,d+1}, ..., u_{1,2d}$};
	\node[right=2pt of t2] (dots) {$\cdots$};
	\node[inputNode,right=2pt of dots] (tM) {$\tiny u_{1,Md-d+1}, ..., u_{1,Md}$};
	\node[output,right=20pt of tM] (path1) {$\textmd{path}_1$};
	\node[output,right=40pt of path1] (payoff1) {$\textmd{payoff}_1$};

	\node[draw, thick, minimum width=38pt, minimum height=114pt] at (-2.45,-1.75) {};

	\draw[stateTransition] (tM) --  (path1);
	\draw[stateTransition] (path1) --  (payoff1);
\node[hiddenB] (thread2) at (-10,-0.5) {Thread B};

\node[inputNode, right=50pt of thread2.west] (t1) {$\tiny u_{2,1} , u_{2,2}, ..., u_{2,d}$};
\node[inputNode, right=2pt of t1] (t2) {$\tiny u_{2,d+1}, ..., u_{2,2d}$};
\node[right=2pt of t2] (dots) {$\cdots$};
\node[inputNode,right=2pt of dots] (tM) {$\tiny u_{2,Md-d+1}, ..., u_{2,Md}$};
\node[output,right=20pt of tM] (path2) {$\textmd{path}_2$};
\draw[stateTransition] (tM) --  (path2);

\node[hiddenC] (thread3) at (-10,-1) {Thread C};

\node[inputNode, right=50pt of thread3.west] (t1) {$\tiny u_{3,1} , u_{3,2}, ..., u_{3,d}$};
\node[inputNode, right=2pt of t1] (t2) {$\tiny u_{3,d+1}, ..., u_{3,2d}$};
\node[right=2pt of t2] (dots) {$\cdots$};
\node[inputNode,right=2pt of dots] (tM) {$\tiny u_{3,Md-d+1}, ..., u_{3,Md}$};
\node[output,right=20pt of tM] (path3) {$\textmd{path}_3$};
\node[output,right=40pt of path2] (payoff3) {$\textmd{payoff}_3$};
\node[output,right=40pt of path3] (payoff2) {$\textmd{payoff}_2$};

\draw[stateTransition] (tM) --  (path3);
\draw[stateTransition] (path2) --  (payoff2);
\draw[stateTransition] (path3) --  (payoff3);

\node[hiddenC] (thread4) at (-10,-1.5) {Thread C};

\node[inputNode, right=50pt of thread4.west] (t1) {$\tiny u_{4,1} , u_{4,2}, ..., u_{4,d}$};
\node[inputNode, right=2pt of t1] (t2) {$\tiny u_{4,d+1}, ..., u_{4,2d}$};
\node[right=2pt of t2] (dots) {$\cdots$};
\node[inputNode,right=2pt of dots] (tM) {$\tiny u_{4,Md-d+1}, ..., u_{4,Md}$};
\node[output,right=20pt of tM] (path4) {$\textmd{path}_4$};
\draw[stateTransition] (tM) --  (path4);

\node[hiddenA] (thread4) at (-10,-2) {Thread A};

\node[inputNode, right=50pt of thread4.west] (t1) {$\tiny u_{5,1} , u_{5,2}, ..., u_{5,d}$};
\node[inputNode, right=2pt of t1] (t2) {$\tiny u_{5,d+1}, ..., u_{5,2d}$};
\node[right=2pt of t2] (dots) {$\cdots$};
\node[inputNode,right=2pt of dots] (tM) {$\tiny u_{5,Md-d+1}, ..., u_{5,Md}$};
\node[output,right=20pt of tM] (path5) {$\textmd{path}_5$};

\node[output,right=40pt of path5] (payoff4) {$\textmd{payoff}_4$};
\draw[stateTransition] (path4) --  (payoff4);

\node[output,right=40pt of path4] (payoff5) {$\textmd{payoff}_5$};
\draw[stateTransition] (tM) --  (path5);
\draw[stateTransition] (path5) --  (payoff5);

\node[hiddenB] (thread5) at (-10,-2.5) {Thread B};

\node[inputNode, right=50pt of thread5.west] (t1) {$\tiny u_{6,1} , u_{6,2}, ..., u_{6,d}$};
\node[inputNode, right=2pt of t1] (t2) {$\tiny u_{6,d+1}, ..., u_{6,2d}$};
\node[right=2pt of t2] (dots) {$\cdots$};
\node[inputNode,right=2pt of dots] (tM) {$\tiny u_{6,Md-d+1}, ..., u_{6,Md}$};
\node[output,right=20pt of tM] (path1) {$\textmd{path}_6$};
\node[output,right=40pt of path1] (payoff1) {$\textmd{payoff}_6$};
\draw[stateTransition] (tM) --  (path1);
\draw[stateTransition] (path1) --  (payoff1);

\node[hiddenB] (thread7) at (-10,-3) {Thread B};

\node[inputNode, right=50pt of thread7.west] (t1) {$\tiny u_{7,1} , u_{7,2}, ..., u_{7,d}$};
\node[inputNode, right=2pt of t1] (t2) {$\tiny u_{7,d+1}, ..., u_{7,2d}$};
\node[right=2pt of t2] (dots) {$\cdots$};
\node[inputNode,right=2pt of dots] (tM) {$\tiny u_{7,Md-d+1}, ..., u_{7,Md}$};
\node[output,right=20pt of tM] (path1) {$\textmd{path}_7$};
\node[output,right=40pt of path1] (payoff1) {$\textmd{payoff}_7$};
\draw[stateTransition] (tM) --  (path1);
\draw[stateTransition] (path1) --  (payoff1);

\node[hiddenA] (thread8) at (-10,-3.5) {Thread A};

\node[inputNode, right=50pt of thread8.west] (t1) {$\tiny u_{8,1} , u_{8,2}, ..., u_{8,d}$};
\node[inputNode, right=2pt of t1] (t2) {$\tiny u_{8,d+1}, ..., u_{8,2d}$};
\node[right=2pt of t2] (dots) {$\cdots$};
\node[inputNode,right=2pt of dots] (tM) {$\tiny u_{8,Md-d+1}, ..., u_{8,Md}$};
\node[output,right=20pt of tM] (path1) {$\textmd{path}_8$};
\node[output,right=40pt of path1] (payoff1) {$\textmd{payoff}_8$};
\draw[stateTransition] (tM) --  (path1);
\draw[stateTransition] (path1) --  (payoff1);
\node[above=10pt of thread1.west] at (-2.35,0) {Statistics};

\end{tikzpicture}
\caption{Second approach using 3 threads and 8 paths. $u_{i,j}$ is the ${(i-1)Md+j}$-th random number or the $j$-th coordinate of the $i$-th $Md$-dimensional number of a low discrepancy sequence. The threads complete their work concurrently, in no particular order.\label{fig:second_approach}}
\end{figure}
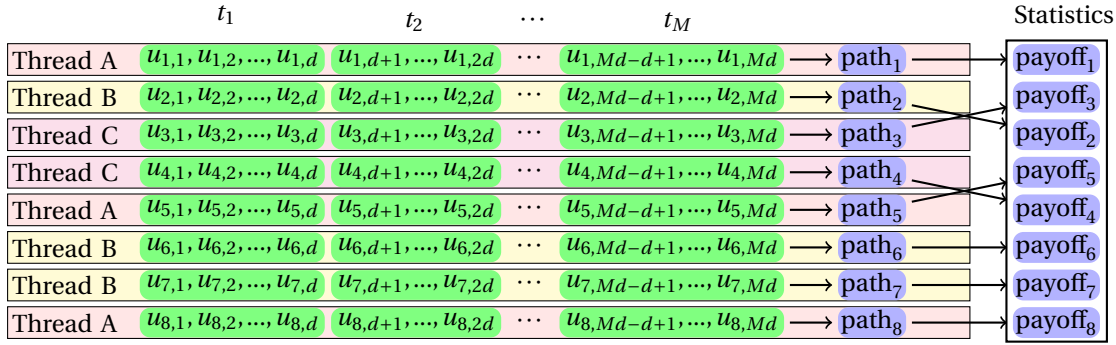

A second approach is to compute in the same thread the path and the random numbers necessary for this path (Figure \ref{fig:second_approach}).
 The random number generator must then be able to skip-ahead, in order to start at $(k-1)  D + 1$ for the $k$-th path, for $k \in \left\{1,..,N\right\}$. Paths will be constructed in a random order, as soon as some CPU time is available. The popular Mersenne-Twister \citep{matsumoto1998mersenne} number generator is then not the best suited for this task, since skipping a medium-sized sequence (typically up to 100,000 numbers) is often slower than generating all those numbers, and skipping is relatively slow even with the best algorithms available \citep{haramoto2008efficient,panneton2006improved}. Instead, combined multiple recursive generator MRG32k3a or MRG63k3a are more appropriate \citep{lecuyer1999good}. Counter based random-number generators such as Philox or Chacha or AES are several orders of magnitude faster since skipping has essentially the same cost as generating one number \citep{salmon2011parallel}. Low discrepancy sequences such as  \citet{sobol1994primer} also allow for a very fast skip-ahead to an arbitrary position.
In practice, it is more efficient to process several paths together, in the same work unit. The logic stays however fundamentally the same.

A third approach is to compute the paths with the random numbers in vectorized manner, time-step by time-step. The $k$-th time-step will query for the $N d$ random numbers starting at $(k-1) N d + 1$. At each time-step all paths are generated, but only for a single point in time. The payoff is then evaluated at its relevant points in time and some $N$-dimensional state variables may be used to keep track of path-dependent features of the payoff. This typically requires more memory than the second approach, as typically, $N \gg D$, but is usually acceptable. Note that, in this approach, because the time-steps are processed sequentially, it is also possible to sample the random numbers from $M$ sub-streams (a sub-stream typically corresponds to a sequence of random numbers starting far away, for example by skipping $2^{64}$ points from the initial position), which may be faster than skipping to $(k-1) N d + 1$ for some random number generators.
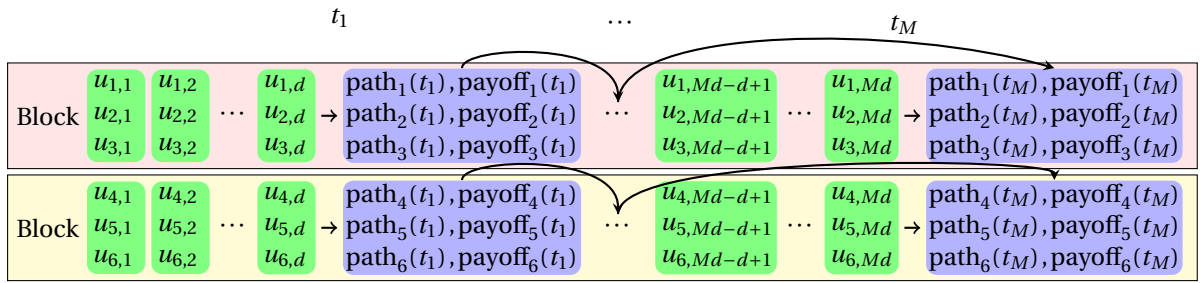
\begin{figure}[h!]
	\begin{tikzpicture}
	\node[hiddenA, minimum height=40pt, minimum width=450pt,anchor=north] (thread1) at (-10,-0) {};
	\node[right=0pt of thread1.west] {Block};
\node[above=10pt of thread1.west] (t1l) at (-13.5,0) {$t_1$};
		\node[] (dotsl) at (-9.75,0.5)  {$\cdots$};
		\node[] (tMl) at (-6, 0.5) {$t_M$};
		\node[inputNode, right=30pt of thread1.west, text width=18pt, anchor=west] (t11) {$\tiny u_{1,1}$\\$\tiny u_{2,1}$\\$\tiny u_{3,1}$};
		\node[inputNode, right=2pt of t11, text width=18pt] (t12) {$\tiny u_{1,2}$\\$\tiny u_{2,2}$\\$\tiny u_{3,2}$};
	\node[right=0pt of t12] (dots) {$\cdots$};
		\node[inputNode, right=0pt of dots, text width=18pt] (t1d) {$\tiny u_{1,d}$\\$\tiny u_{2,d}$\\$\tiny u_{3,d}$};
\node[output,right=10pt of t1d, text width=88pt] (path1) {$\textmd{path}_1(t_1)\,,\textmd{payoff}_1(t_1)$\\ $\textmd{path}_2(t_1)\,,\textmd{payoff}_2(t_1)$\\$\textmd{path}_3(t_1)\,,\textmd{payoff}_3(t_1)$ };
\draw[stateTransition] (t1d) --  (path1);
	\node[right=5pt of path1 ] (dotsm) {$\cdots$};
	\node[above=5pt of path1] (above1) {};
	\node[above=20pt of dotsm] (abovedots) {};
	
	\draw[-stealth, thick](path1.north) ..controls(above1) and (abovedots) ..  (dotsm.north){};
	\node[inputNode, right=5pt of dotsm, text width=42pt] (t11) {$\tiny u_{1,Md-d+1}$\\$\tiny u_{2,Md-d+1}$\\$\tiny u_{3,Md-d+1}$};
	\node[right=0pt of t11] (dots) {$\cdots$};
	\node[inputNode, right=0pt of dots, text width=24pt] (t1d) {$\tiny u_{1,Md}$\\$\tiny u_{2,Md}$\\$\tiny u_{3,Md}$};
	\node[output,right=10pt of t1d, text width=95pt] (path1) {$\textmd{path}_1(t_M)\,,\textmd{payoff}_1(t_M)$\\ $\textmd{path}_2(t_M)\,,\textmd{payoff}_2(t_M)$\\$\textmd{path}_3(t_M)\,,\textmd{payoff}_3(t_M)$ };
	\draw[stateTransition] (t1d) --  (path1);
	\draw[-stealth, thick](dotsm.north) ..controls(dotsl) and (tMl) ..  (path1.north){};

	\node[hiddenB, minimum height=40pt, minimum width=450pt,below=2pt of thread1] (thread2) {};
\node[right=0pt of thread2.west] {Block};
\node[inputNode, right=30pt of thread2.west, text width=18pt, anchor=west] (t11) {$\tiny u_{4,1}$\\$\tiny u_{5,1}$\\$\tiny u_{6,1}$};
\node[inputNode, right=2pt of t11, text width=18pt] (t12) {$\tiny u_{4,2}$\\$\tiny u_{5,2}$\\$\tiny u_{6,2}$};
\node[right=0pt of t12] (dots) {$\cdots$};
\node[inputNode, right=0pt of dots, text width=18pt] (t1d) {$\tiny u_{4,d}$\\$\tiny u_{5,d}$\\$\tiny u_{6,d}$};
\node[output,right=10pt of t1d, text width=88pt] (path1) {$\textmd{path}_4(t_1)\,,\textmd{payoff}_4(t_1)$\\ $\textmd{path}_5(t_1)\,,\textmd{payoff}_5(t_1)$\\$\textmd{path}_6(t_1)\,,\textmd{payoff}_6(t_1)$ };
\draw[stateTransition] (t1d) --  (path1);
\node[right=5pt of path1 ] (dotsm) {$\cdots$};
\node[above=5pt of path1] (above1) {};
\node[above=20pt of dotsm] (abovedots) {};

\draw[-stealth, thick](path1.north) ..controls(above1) and (abovedots) ..  (dotsm.north){};

\node[inputNode, right=5pt of dotsm, text width=42pt] (t11) {$\tiny u_{4,Md-d+1}$\\$\tiny u_{5,Md-d+1}$\\$\tiny u_{6,Md-d+1}$};
\node[right=0pt of t11] (dots) {$\cdots$};
\node[inputNode, right=0pt of dots, text width=24pt] (t1d) {$\tiny u_{4,Md}$\\$\tiny u_{5,Md}$\\$\tiny u_{6,Md}$};
\node[output,right=10pt of t1d, text width=95pt] (path1) {$\textmd{path}_4(t_M)\,,\textmd{payoff}_4(t_M)$\\ $\textmd{path}_5(t_M)\,,\textmd{payoff}_5(t_M)$\\$\textmd{path}_6(t_M)\,,\textmd{payoff}_6(t_M)$ };
\draw[stateTransition] (t1d) --  (path1);
\node[above=5pt of path1] (above1) {};
\draw[-stealth, thick](dotsm.north) ..controls(abovedots) and (above1) ..  (path1.north){};

\end{tikzpicture}
\caption{Hybrid vectorized approach with two blocks of 3 paths each. $u_{i,j}$ is the ${(i-1)Md+j}$-th random number or the $j$-th coordinate of the $i$-th $Md$-dimensional number of a low discrepancy sequence. \label{fig:hybrid_approach}}
\end{figure}

The most flexible approach is a hybrid approach which consists in applying the third approach, not on the total number of paths, but on a block of paths, where different blocks are processed in different threads (Figure \ref{fig:hybrid_approach}). A similar skipping logic as in the second approach would then apply.

The mean and variance of the Monte-Carlo simulation are calculated from the payoff values on each path. In the second approach or in the hybrid approach, calculating the mean and variance statistics in running manner allows to reduce the memory consumption, as there is no need to keep an array of size $N$ of the payoff values in memory. But, as a consequence, the order of the summations involved will be random, as dictated by the operating system scheduler choices and the number of concurrent threads. In the hybrid approach, another technique would be to store the statistics of each block in an array, whose index follows the blocks order. Assembling the statistics would then consist in summing the blocks sums and variances in the natural order of the array. The latter approach may also suffer from a reproducibility of the end statistics if the size of a block changes\footnote{For example, we may allow larger blocks for some machines with more memory or less CPU cores.}.

\section{Modified algorithms}
Algorithm \ref{alg:naive_kahan} is the pseudo-code for naive calculation of the mean and variance, using the correction of Kahan for the summation. 
\begin{algorithm}
	\caption{Naive algorithm with Kahan summation}\label{alg:naive_kahan}
\begin{algorithmic}[1]\onehalfspacing
\STATE	$S \gets 0$; $T \gets 0$; $S^\star \gets 0$; $T^\star \gets 0$; $k \gets 0$
\WHILE {$k < n$}
\STATE $ k \gets k + 1$
\STATE $y \gets x_k - S^\star$; $t \gets S + y$
\STATE $S^\star \gets  (t-S) - y$
\STATE $S \gets t$
\STATE $y \gets x_k^2 - T^\star$; $t \gets T + y$
\STATE $T^\star \gets (t- T) - y$
\STATE $T \gets t$
\ENDWHILE
\STATE $M \gets  S/n$; $V \gets T/n - M^2$
\end{algorithmic}
\end{algorithm}

Algorithm \ref{alg:naive_shifted_kahan} uses the first observation as constant shift for the calculation of the variance, in order to avoid catastrophic cancellation errors.
\begin{algorithm}
	\caption{Shifted naive algorithm with Kahan summation}\label{alg:naive_shifted_kahan}
	\begin{algorithmic}[1]\onehalfspacing
		\STATE	$S \gets 0$; $T \gets 0$; $S^\star \gets 0$; $T^\star \gets 0$; $k \gets 0$
		\WHILE {$k < n$}
		\STATE $ k \gets k + 1$
		\IF{$k = 1$} 
	\STATE{$ K \gets  x_k$}
	\ENDIF
		\STATE $y \gets x_k - K - S^\star$; $t \gets S + y$
		\STATE $S^\star \gets  (t-S) - y$
		\STATE $S \gets t$
		\STATE $y \gets (x_k-K)^2 - T^\star$; $t \gets T + y$
		\STATE $T^\star \gets (t- T) - y$
		\STATE $T \gets t$
		\ENDWHILE
		\STATE $M \gets  S/n+K$; $V \gets T/n - (M-K)^2$
	\end{algorithmic}
\end{algorithm}

Algorithm \ref{alg:chan_kahan} is the pseudo-code for the algorithm recommended by \citet{chan1979computing} with the correction of Kahan. The mean is computed through the naive algorithm (Equation \ref{eqn:naive_mean}) and the variance with the refined algorithm (Equation \ref{eqn:chan_var}).  
\begin{algorithm}
	\caption{Chan and Lewis algorithm with Kahan summation}\label{alg:chan_kahan}
	\begin{algorithmic}[1]\onehalfspacing
		\STATE	$S \gets 0$; $T \gets 0$; $M \gets 0$;  $S^\star \gets 0$; $T^\star \gets 0$; $k \gets 0$
		\WHILE {$k < n$}
\STATE $ k \gets k + 1$
		\STATE $z \gets  (k-1) (x_k - M)^2/k$
	\STATE $y \gets z - T^\star$; $t \gets T + y$
\STATE $T^\star \gets (t- T) - y$
\STATE $T \gets t$
		\STATE $y \gets x_k - S^\star$; $t \gets S + y$
		\STATE $S^\star = (t-S) - y$
		\STATE $S \gets t$
		\STATE $M \gets  S/k$;
		\ENDWHILE
		\STATE   $V\gets T/n$
	\end{algorithmic}
\end{algorithm}
Algorithm \ref{alg:ling_kahan} is the pseudo-code for the algorithm recommended by \citet{ling1974comparison} with the correction of Kahan. The difference with the previous algorithm is the calculation of the mean, which relies on Equation \ref{eqn:chan_mean}.
\begin{algorithm}
	\caption{Ling algorithm with Kahan summation}\label{alg:ling_kahan}
	\begin{algorithmic}[1]\onehalfspacing
		\STATE	$M \gets 0$; $T \gets 0$; $M^\star \gets 0$; $T^\star \gets 0$; $k \gets 0$
		\WHILE {$k < n$}
		\STATE $ k \gets k + 1$
		\STATE $z \gets  (k-1) (x_k - M)^2/k$
		\STATE $y \gets z - T^\star$; $t \gets T + y$
		\STATE $T^\star \gets (t- T) - y$
		\STATE $T \gets t$
		\STATE $z \gets  (x_k - M)/k$
		\STATE $y \gets z - M^\star$; $t \gets M + y$
		\STATE $M^\star = (t-M) - y$
		\STATE $M \gets t$
		\ENDWHILE
		\STATE $V\gets T/n$
	\end{algorithmic}
\end{algorithm}
This last algorithm could also be implemented using the formulation presented in \citep{knuth2014art}. Doing so does not change the results presented in this paper.

\section{Examples}
\subsection{Normal distribution}
The first example samples 100,000 random numbers of the normal distribution $\mathcal{N}(\mu, \sigma)$ with a large mean $\mu=100000$ compared to the deviation $\sigma=1$ similarly to the experiments in \citep{ling1974comparison}. The difference is that we use 64-bit floating point number arithmetic, and thus a larger mean is necessary to put in evidence some of the differences between the algorithms. The average error in $M$ and $V$ on 100 runs is reported in Table \ref{tbl:normal_error}. The reference values are obtained by an exact multiple precision implementation. We report the error using the raw random samples, as well as with the samples sorted in ascending order. The sorting is used to simulate the multi-threading effect: as explained in Section \ref{sec:more_threads}, a parallel calculation would use the samples in a different order than the original raw random sequence. Here, our choice of different order is sorted numbers.

\begin{table}[h]\caption{We report the relative absolute error on the original random sequence of mrg63k3a as well as on the reordered samples in ascending order. The  average reference values are $M=99999.99994407306$ and $V= 1.0012459250239656$. \label{tbl:normal_error}}
	\centering{
	\begin{tabular}{lcccc}\toprule
		Algorithm  &  \multicolumn{2}{c}{Unordered} & \multicolumn{2}{c}{Sorted ascending}\\\cmidrule(lr){2-3} \cmidrule(lr){4-5}
		& error in $M$ & error $V$ 	& error in $M$ & error $V$ \\ \midrule
		Naive & 6.92E-15 & 1.50E-04 & 6.87E-15 & 1.56E-04\\
		Kahan correction & 5.11E-17 & 1.32E-06 & 5.11E-17 & 1.32E-06 \\
		Klein correction & 5.11E-17 & 1.32E-06 & 5.11E-17 & 1.32E-06 \\
		Shifted Naive with Kahan correction & 0 & 1.17E-16 & 0 & 2.52E-15 \\
		Ling & 6.11E-15 & 2.33E-12&  6.50E-15& 6.36E-10\\
		Ling with Kahan correction & 0 & 1.92E-14 & 0 & 2.21E-14\\
		Chan and Lewis with Kahan correction & 5.11E-17 & 3.16E-14&  5.11E-17 & 3.07E-14\\
\bottomrule
	\end{tabular}}
	\end{table}

When calculated without correction term for the summation, the mean value calculated is slightly inaccurate compared to the exact implementation. Furthermore, its value changes when the samples are ordered differently. This is not the case when Kahan or Klein summation corrections are included, and actually we do not measure any error\footnote{in double machine accuracy.} on this test with the corrections. The correction also helps achieving a slightly better accuracy for the variance. The refined algorithm (Equation \ref{eqn:chan_var}) improves its accuracy much further.
In this test, the most accurate algorithm is the shifted naive algorithm with Kahan correction (Algorithm \ref{alg:naive_shifted_kahan}).

\subsection{Uniform distribution under 32-bit precision}

In the second example, we evaluate the accuracy of the Klein and Kahan corrections using 32-bit floating point numbers. We adapt the example of \citet{klein2006generalized}, summing 50 million random numbers uniformly distributed in the interval (0,1), but instead of simply computing the sum, we also compute the mean value and variance (using the 32-bit floating point representation).

\begin{table}[h]\caption{We report the absolute error on the original random sequence of mrg63k3a as well as on the reordered samples in ascending order. The  average reference values are $S=2.4999470327539172E7$, $M=0.49998940655078344$ and $V=0.08334439919010794$. \label{tbl:uniform_error}}
	\centering{
		\begin{tabular}{lcccccc}\toprule
			Algorithm  &  \multicolumn{3}{c}{Unordered} & \multicolumn{3}{c}{Sorted ascending}\\\cmidrule(lr){2-4} \cmidrule(lr){5-7}
			& in $S$ & in $M$ & in $V$ & in $S$	&  in $M$ & in $V$ \\ \midrule
			Naive & 8.22E+6  & 3.29E-1 & 1.35E-0 & 8.22E+6 & 3.29E-1 & 1.68E-0 \\
			Kahan correction & 3.28E-1 & 3.24E-8  & 8.87E-9 & 3.28E-1 & 3.24E-8  & 8.87E-9 \\
			Klein correction & 3.28E-1 & 3.24E-8  & 9.65E-6 & 2.64E+4 & 1.06E-4 & 6.32E-4 \\
			Knuth correction & 1.38E+2 & 5.52E-6 & 5.95E-3 & 3.29E+6  & 1.32E-2 &8.22E-2  \\
			Shifted Naive with Kahan correction & 3.28E-1 & 3.24E-8  & 8.87E-9 & 3.28E-1  & 3.24E-8  & 8.87E-9\\
			Ling & 3.84E+4 & 1.53E-4& 7.43E-2 & 1.23E+7 & 4.90E-1 & 7.15E-2\\
			Ling with Kahan correction &  1.67E+0 & 2.73E-8 & 8.87E-9 &  1.67E+0 & 2.73E-8 & 8.87E-9\\
			Chan and Lewis with Kahan correction & 3.28E-1 & 3.24E-8 & 8.87E-9  & 3.28E-1 & 3.24E-8 & 8.87E-9\\
			\bottomrule
	\end{tabular}}
\end{table}

\begin{figure}[h]
	\begin{center}
		\subfloat[][The 7-th sequence shows a small difference in favor of Kahan. Kahan is as accurate on the ordered sequences.]{\includegraphics[width=0.5\textwidth]{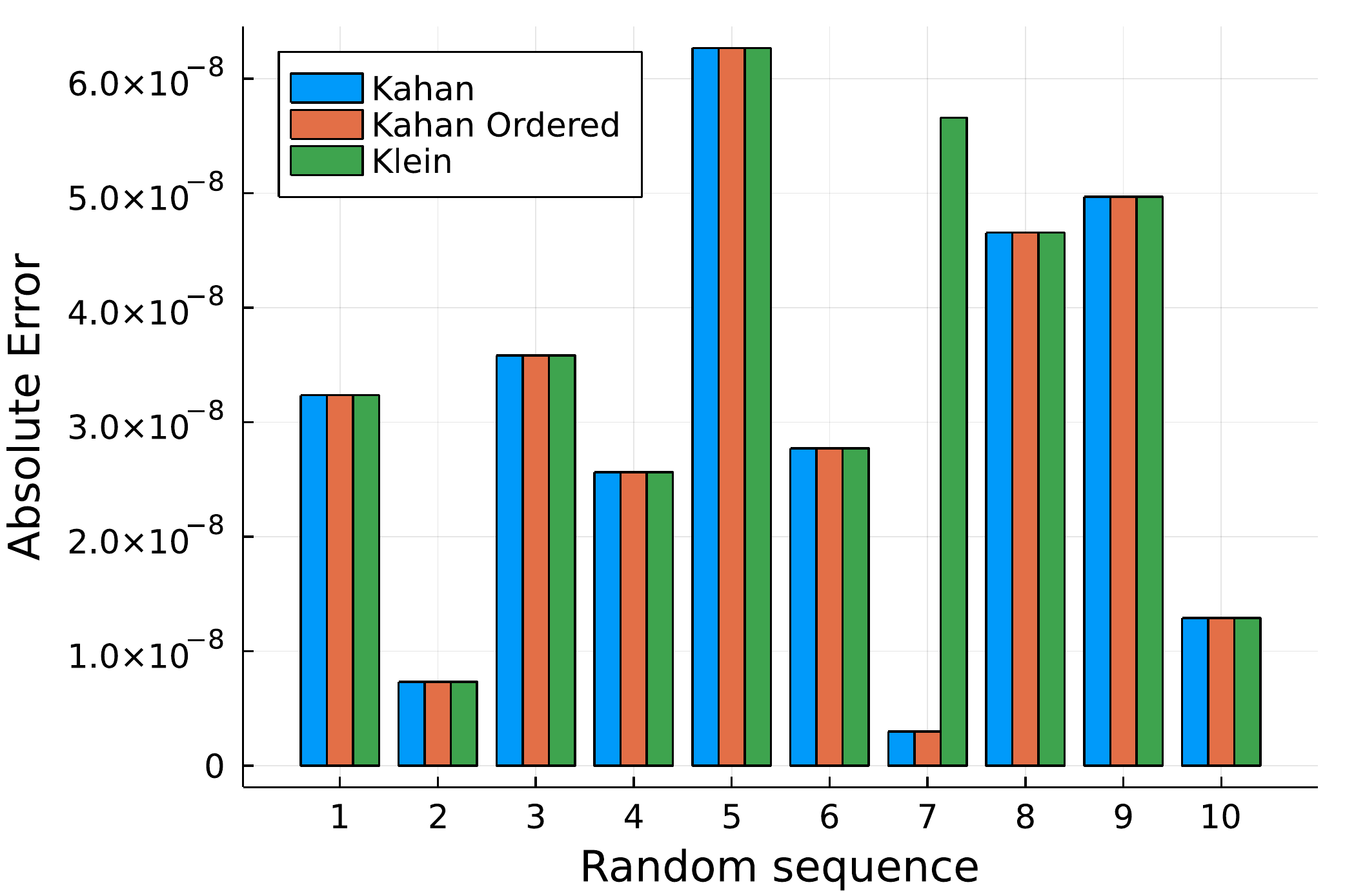}}
		\subfloat[][Klein correction is much less accurate on the ordered sequences.]{\includegraphics[width=0.5\textwidth]{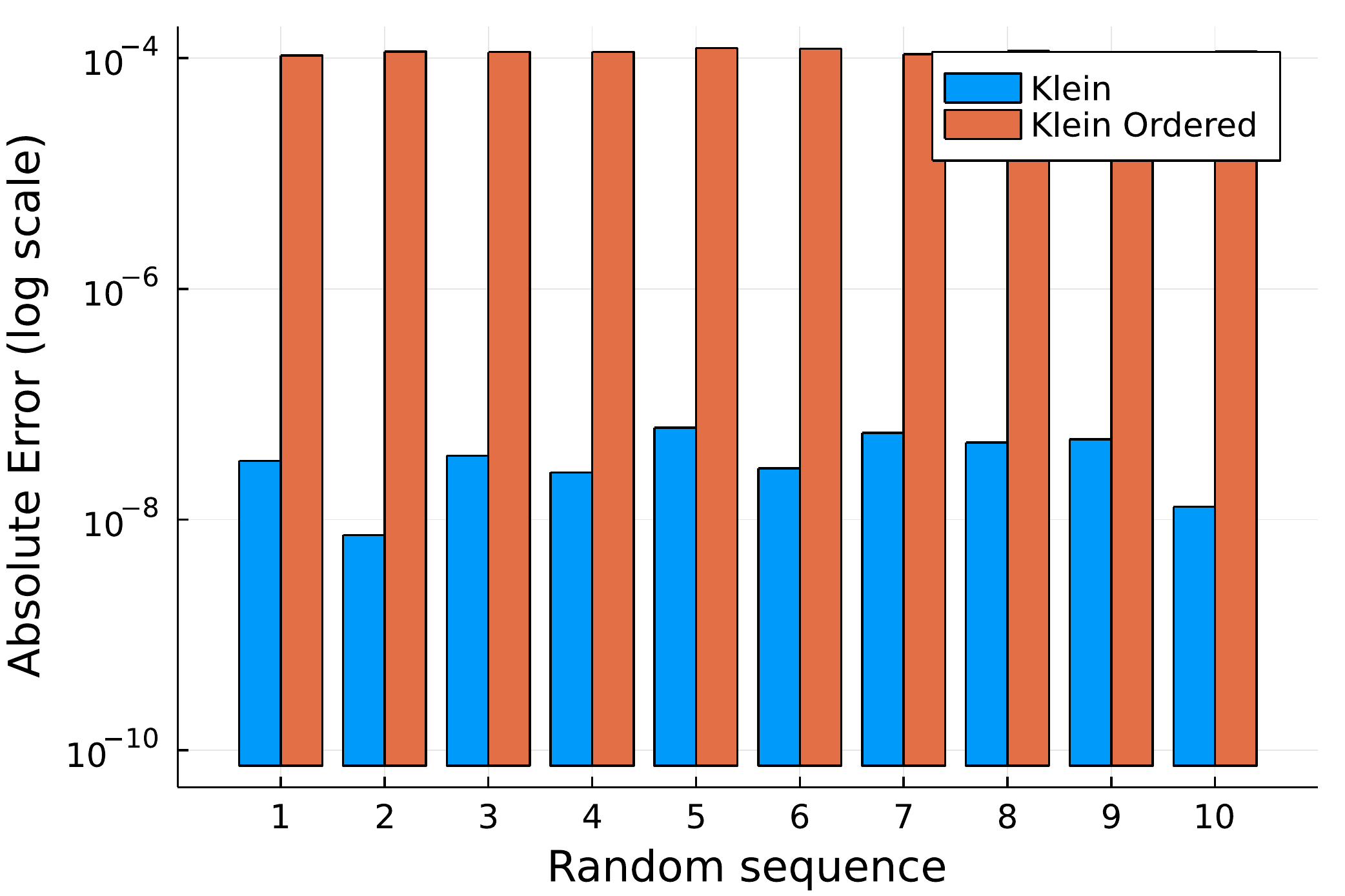}}
	\end{center}
	\caption{Absolute errors on 10 random sequences of 50 million numbers in 32-bit arithmetic. \label{fig:kahan_klein_32}}
\end{figure}

\citet{klein2006generalized} finds that their second-order iterative algorithm is the most accurate to compute the sum. Our test does not reproduce this fact: in Table \ref{tbl:uniform_error}, the algorithm\footnote{The second-order iterative algorithm is given in pseudo-code in \citep[p. 283]{klein2006generalized}} is not more accurate than the Kahan correction. Furthermore, we find that the second-order iterative algorithm of \citet{klein2006generalized} is significantly less accurate on the ordered set, likely because the algorithm corrects the sum at the end instead of at every step. In contrast, the Kahan correction leads to the same results on both sets, and is accurate up to the 32-bit machine epsilon. Using longer (100 million) or shorter (30 million) sequences or different random sequences lead to the same observations (Figure \ref{fig:kahan_klein_32}).
The choice of 50 million is not innocuous: it is not very far from the inverse of the 32-bit machine epsilon. With 64-bit floating point numbers, the threshold to see an effect of the summation correction would be much higher.

\subsection{Monte-Carlo simulations of financial derivative contracts}
The third example consists in pricing a financial derivative contract under the Black-Scholes model. We choose a European binary asset or nothing option of strike price $K=1.5$ and maturity $T=1$ with a quantity $q=10^6$. The payoff at expiry of such an option is simply $V(S, T)=1_{S(T) \geq K} q S(T)$. The Black-Scholes model parameters are a volatility $\sigma=50\%$ and an asset initial price $S(0)=1$. We use a single-step Monte-Carlo simulation to price such a contract and computes its $\Gamma$ sensitivity using the formula \begin{equation}	\Gamma=\frac{V(S+\epsilon S,0)-2V(S,0)+V(S-\epsilon S,0)}{S^2 \epsilon^2}\,,
\end{equation}
where $\epsilon = 0.01$ and $V(S,0)$ is the mean value of the Monte-Carlo simulation computed over 1,000,000 paths for the option contract specified. We reuse the same random sequence to calculate $V(S+\epsilon S,0)$ and $V(S-\epsilon S,0)$.
Because of the discontinuity in the payoff at maturity, even with such a large number of paths, the estimate of the $\Gamma$ will not be very good. We however do not intend to compute the $\Gamma$ value accurately,  we merely want to measure the accuracy of the algorithm used for the mean value calculation and its impact on the finite difference estimate. Our concern is reproducibility. This example is representative for more exotic products and more complex models involving more dimensions in the Monte-Carlo simulation.

%

\begin{table}[h]\caption{We report the mean absolute error on the original random sequence of mrg63k3a as well as on the reordered samples in ascending order for an asset or nothing option over 10 runs. The reference values of the first run are $M=288704.05831425334$ and $V=$ 5.2979360217605115E11   and $\Gamma=1244141.9891041005$. Reference values are calculated using multiple precision arithmetic. The theoretical value for $\Gamma$ is   644002.9482875252. \label{tbl:assetornothing_error}}
	\centering{
		\begin{tabular}{lcccccc}\toprule
			Algorithm  &  \multicolumn{3}{c}{Unordered} & \multicolumn{3}{c}{Sorted ascending}\\\cmidrule(lr){2-4} \cmidrule(lr){5-7}
			& error in $M$ & error $V$ 	&  error in $\Gamma$ & error in $M$ & error $V$	&  error in $\Gamma$  \\ \midrule
			Naive & 4.26E-9 & 8.38E-3 & 4.91E-5 & 1.13E-9 & 6.67E-3 & 3.13E-5 \\
			Kahan correction & 2.91E-11 & 6.71E-5 & 6.40E-7 &  2.91E-11 & 6.71E-5 & 6.40E-7  \\
			Shifted Naive with Kahan correction & 2.91E-11 & 1.28E-4 & 6.40E-7 & 2.91E-11 & 6.71E-5 & 6.40E-7 \\
			Ling & 7.98E-9 & 9.45E-3&  2.07E-4& 1.87E-9 & 4.96E-3 & 3.91E-5\\
			Ling with Kahan correction & 0 &  1.22E-5 & 0 & 0 &  1.22E-5 & 0 \\
			Chan and Lewis with Kahan correction & 2.91E-11 & 1.22E-5 & 6.40E-7 &  2.91E-11 & 1.22E-5 & 6.40E-7  \\
			\bottomrule
	\end{tabular}}
\end{table}
Table \ref{tbl:assetornothing_error} shows that $\Gamma$ estimate is less accurate and depends on the order of summation for the naive and refined algorithms. In particular, the refined algorithm is not more accurate than the naive one. This is also true for the variance. 

When the Kahan or Klein summation correction is added, we end up with price estimates with an accuracy below machine epsilon on the all 10 runs (no loss of accuracy). The error in the estimate of $\Gamma$ is close to machine epsilon limits: a relative change in price of machine epsilon leads to a change in $Gamma$ of around $10^{-12}$ due to the scaling by $\epsilon^2$. The error in the variance calculation stays below machine epsilon in relative value (the actual variance is very large), except for the shifted naive algorithm, which exhibits some variation between the ordered and unordered sets. The relative change in variance is however only around twice the machine epsilon. The other algorithms with correction (including the naive one) lead to the exact same statistics (price, $\Gamma$, variance) on the ordered and raw sets: the results are exactly reproducible regardless of the order of summation.

The fourth example is the equivalent binary cash or nothing option, with payoff at maturity  $V(S, T)=1_{S(T) \geq K} q$. In this case, the mean is calculation involves the same numbers $q$, albeit not necessarily ordered the same way or with the same frequency. We use a total of 10 million paths, and a single run.
\begin{table}[h]\caption{We report the absolute error on the original random sequence of mrg63k3a as well as on the reordered samples in ascending order for a binary cash or nothing option for a single run. The reference values are $M=144533.5$ and $V=$ 1.2364357974210797E11 and $\Gamma= 658999.9999999418$. The theoretical value for $\Gamma$ is  509875.13993584045. \label{tbl:cashnothing_error}}
	\centering{
		\begin{tabular}{lcccccc}\toprule
			Algorithm  &  \multicolumn{3}{c}{Unordered} & \multicolumn{3}{c}{Sorted ascending}\\\cmidrule(lr){2-4} \cmidrule(lr){5-7}
			& error in $M$ & error $V$ 	&  error in $\Gamma$ & error in $M$ & error $V$	&  error in $\Gamma$  \\ \midrule
			Naive & 0 &0 & 0 &0 &0&0 \\
		    Ling  & 1.32E-8 & 1.22E-4&  6.43E-5& 3.46E-9 & 2.70E-3 & 4.69E-5\\
			Ling with Kahan correction & 0 & 0& 0 & 0 & 0& 0 \\
			Chan and Lewis with Kahan correction & 0 & 0& 0 & 0 & 0& 0 \\\bottomrule
	\end{tabular}}
\end{table}
It is expected then, that the naive algorithm will lead to an exact result, while the refined algorithm does not, because, in the refined algorithm, the sum involves a division by $k+1$ within the iteration, which is not exact in the floating point representation. This is confirmed in Table \ref{tbl:cashnothing_error}. The sum correction algorithms allow to solve this issue on this example. But in general, this example may serve as warning that the refined algorithm is highly dependent towards the ordering of the samples even in seemingly trivial cases. 

If we let the binary cash option pay a small rebate of $r=0.01$ such that  $V(S, T)=1_{S(T) \geq K} q + 1_{S(T) < K} r$ then the calculation of the mean will involve averaging over samples of value $q$ and samples of value $r$. In this situation, the refined algorithm will perform better, as expected, than the naive algorithm, without summation correction.

In practice, the naive calculation of the mean (Equation \ref{eqn:naive_mean}) is no less accurate than the slightly more refined calculation of the mean (Equation \ref{eqn:chan_mean}) when the Kahan or Klein corrections are included: the errors reported are below the double floating point representation accuracy. The results of the first example however show that the refined calculation of the variance is more accurate and the sum corrections allow to increase its accuracy further. The examples above do not allow to distinguish between the two kinds of summation correction.

\section{Conclusion}
The refined algorithm for the mean and variance recommended in \citep{chan1979computing} is improved in terms of accuracy and sensitivity of the result to the order of the data, by using the robust summation of \citet{kahan1965further} for the two sums involved. The additional complexity and cost is negligible in practical applications.

The  \citet{kahan1965further} summation alone is found to be adequate for the calculation of the mean in a naive manner, but not for the variance. The compensating approach used in \cite{chan1979computing} to calculate the mean slightly improves the accuracy, but is in general not necessary when a summation correction is included. 

The addition of a constant shift (taking the first observation as shift) to the naive algorithm solves the catastrophic cancellation error that the original algorithm exhibits for the calculation of the variance, but leads to non reproducible estimates of the variance. This however may not be a practical issue for two reasons: the mean is usually the relevant statistic in terms of reproducibility, and the relative difference in variance is of the order of machine epsilon.

The second-order summation correction of \citet{klein2006generalized} is found to be less accurate than the simpler Kahan correction. Although the difference may not matter as much in 64-bit floating point arithmetic, it is very visible in 32-bit floating point arithmetic.

\funding{This research received no external funding.}
\conflictsofinterest{The authors declare no conflict of interest.}
\supplementary{Source code illustrating the examples in this paper is available at \url{https://github.com/jherekhealy/MonteCarloMeanVarianceExamples.jl}}
\externalbibliography{yes}
\bibliography{mean_variance.bib}
\appendixtitles{no}
\end{document}